\documentclass[12pt]{article}
\usepackage[dvips]{graphicx}
\usepackage{epsfig}
\usepackage{latexsym}
\usepackage{amssymb}
\usepackage{amsmath}
\usepackage{wrapfig}
\usepackage{boxedminipage}
\usepackage{epsfig}
\usepackage{setspace}
\usepackage{subfigure}
\usepackage{hyperref}
\newcommand{\reef}[1]{(\ref{#1})}

\DeclareSymbolFont{AMSb}{U}{msb}{m}{n}
\DeclareMathSymbol{\IN}{\mathbin}{AMSb}{"4E}
\DeclareMathSymbol{\IZ}{\mathbin}{AMSb}{"5A}
\DeclareMathSymbol{\IR}{\mathbin}{AMSb}{"52}
\DeclareMathSymbol{\Q}{\mathbin}{AMSb}{"51}
\DeclareMathSymbol{\II}{\mathbin}{AMSb}{"49}
\DeclareMathSymbol{\IC}{\mathbin}{AMSb}{"43}
\DeclareMathSymbol{\IP}{\mathbin}{AMSb}{"50}
\DeclareMathSymbol{\IH}{\mathbin}{AMSb}{"48}
\DeclareMathSymbol\IA{\mathalpha}{AMSb}{"41}
\DeclareMathSymbol\IS{\mathalpha}{AMSb}{"53}

\def\Q{{\cal Q}}

\begin{document}

\begin{flushright}
%USC-04-05%\\
%DCPT--04/??
\end{flushright}
\begin{center} {\Large \bf A Topology--Changing First Order Phase Transition}
\bigskip\bigskip

{\Large \bf and the}

\bigskip\bigskip

{\Large \bf   Dynamics of Flavour}

\end{center}

\bigskip \bigskip \bigskip

\centerline{\bf Tameem  Albash, Veselin Filev, Clifford V. Johnson\footnote{Also, Visiting Professor at the Centre for Particle Theory, Department of Mathematical Sciencs, University of Durham, Durham DH1  3LE,  U.K.}, Arnab Kundu}

\bigskip
\bigskip

  \centerline{\it Department of Physics and Astronomy }
\centerline{\it University of
Southern California}
\centerline{\it Los Angeles, CA 90089-0484, U.S.A.}

\bigskip
\bigskip

\centerline{\small \tt albash, filev, johnson1,  akundu  @usc.edu}

\bigskip
\bigskip

%\maketitle

\begin{abstract}
  In studying the dynamics of large $N_c$, $SU(N_c)$ gauge theory at
  finite temperature with fundamental quark flavours in the quenched
  approximation, we observe a first order phase transition. A quark
  condensate forms at finite quark mass, and the value of the
  condensate varies smoothly with the quark mass for generic regions
  in parameter space.  At a particular value of the quark mass, there
  is a finite discontinuity in the condensate's vacuum expectation
  value, corresponding to a first order phase transition. We study the
  gauge theory via its string dual formulation using the AdS/CFT
  conjecture, the string dual being the near--horizon geometry of
  $N_c$ D3--branes at finite temperature,
  AdS$_5$--Schwarzschild$\times S^5$, probed by a D7--brane. The
  D7--brane has topology $\IR^4\times S^3\times S^1$ and allowed
  solutions correspond to either the $S^3$ or the $S^1$ shrinking away
  in the interior of the geometry. The phase transition represents a
  jump between branches of solutions having these two distinct
  D--brane topologies. The transition also appears in the meson
  spectrum.

\end{abstract}
\newpage \baselineskip=18pt \setcounter{footnote}{0}

%%%%%%%%%%%%%%%%%%%%%%%%%%%%%%%%%%%%%%%%%%%%%%%%%%%%%%%%%%%%%%%%%%%%%%%%%%%%%%%

\section{Introduction}
\label{sec:introduction}

String theory is a powerful tool for probing the strongly coupled
dynamics of gauge theory, physics which is of vital importance
especially in the context of the strong nuclear interactions.
Gauge/string correspondences have enlarged and refined the toolbox
available for such studies, and there is a large literature on the
subject, with several powerful examples such as the AdS/CFT
correspondence\cite{Maldacena:1997re, Witten:1998qj, Gubser:1998bc}  and deformations
thereof\cite{Aharony:1999t}.

However, we are still some way from describing the ``realistic'' dynamics of
QCD using this approach. The main challenges that remain
include getting access to low $N_c$, fully including dynamical quarks
in the fundamental representation of $SU(N_c)$, and getting reliable
control of the non--supersymmetric regime.

It is to be hoped (if not expected) that even if we cannot obtain a
controllable string dual of QCD, there may be considerable progress to
be made in capturing physical phenomena that are in the same
universality class as those of QCD. This is the motivation of the
present work, which is part of a series of studies upon which we hope
to report new and interesting results.

We study the geometry of AdS$_5$--Schwarzschild $\times S^5$, which is
the decoupled/near--horizon geometry of $N_c$ D3-branes, where $N_c$
is large and set by the (small, for reliability) curvature of the
geometry. The physics of closed type~IIB string theory in this
background is dual to the physics of ${\cal N}=4$ supersymmetric
$SU(N_c)$ gauge theory in four dimensions, with the supersymmetry broken by being at finite
temperature\cite{Witten:1998zw}. The temperature is set by the horizon
radius of the Schwarzschild black hole, as we will recall below.

We introduce a D7--brane probe into the background. Four of the
brane's eight world--volume directions are parallel with those of the
D3--branes, and three of them wrap an $S^3\subset S^5$. The remaining
direction lies in the radial direction of the asymptotically AdS$_5$
geometry.

Such a D3--D7 configuration controls the physics of
the $SU(N_c)$ gauge theory with a dynamical quark in the fundamental
representation\cite{Karch:2002sh}.  The configuration (at zero temperature)
preserves ${\cal N}=2$ supersymmetry in $D=4$, and the quark is part
of a hypermultiplet.  Generically, we will be studying the physics at finite
temperature, so supersymmetry will play no explicit role here.

We are studying the D7--brane as a probe only, corresponding to taking the $N_c \gg N_f$ limit, and therefore there is no backreaction on the background geometry. This is roughly analogous
to the quenched approximation in lattice QCD.  The quark mass and other flavour physics--such as the
vacuum expectation value (vev) of a condensate and the  spectrum of mesons
that can be constructed from the quarks--are all physics which are
therefore invisible in the background geometry. We will learn nothing
new from the background; our study is of the response of the probe D7--branes to the background, and this is where the new physics emerges from.

We carefully study the physics of the probe itself as
it moves in the background geometry. The coordinates of the probe in
the background are fields in an effective D7--brane world--volume
theory, and the geometry of the background enters as couplings controlling the dynamics of
those fields. One such coupling in the effective model represents the
local separation, $L(u)$, of the D7--brane probe from the D3--branes,
where $u$ is the radial AdS$_5$--Schwarzschild coordinate.

In fact, the asymptotic value of the separation between the D3--branes and
D7--brane for large $u$ yields the bare quark mass $m_q$ and the
condensate vacuum expectation value (vev) $ \langle \bar{\psi} \psi
\rangle$ as follows \cite{Polchinski:2000uf} \cite{Kruczenski:2003uq}:
\begin{equation} \label{eqt: L}
\lim_{u \rightarrow \infty} L(u) = m + \frac{c}{u^2} + \dots
\end{equation}
where $m = 2 \pi \alpha' m_q$ and $-c = \langle \bar{\psi} \psi \rangle / (8 \pi^3 \alpha' N_f \tau_{\mathrm{7}})$, where in this paper $N_f = 1$. (The fundamental string tension is defined as $T=1/(2\pi\alpha^\prime)$ here, and the D7--brane tension is $\tau_{\mathrm{7}}=(2\pi)^{-7}(\alpha^\prime)^{-4}$.) The
zero temperature behaviour of the D7--branes in the geometry is
simple. The D7--brane worldvolume actually {\it vanishes} at finite~$u$, corresponding to the part of the brane wrapped on the $S^3$
shrinking to zero size.  The location in $u$ where this vanishing
happens encodes the mass of the quark, or equivalently, the separation
of the probe from the D3--branes.  In addition, in the zero temperature background, the only value of $c$ allowed is zero, meaning no condensate is allowed to form, as is expected from supersymmetry.

The finite temperature physics introduces an important new feature.
As is standard\cite{Gibbons:1979xm}, finite temperature is studied by
Euclideanizing the geometry and identifying the temperature with the
period of the time coordinate.  The horizon of the background geometry
is the place where that $S^1$ shrinks to zero size. The D7--brane
 is also wrapped on this $S^1$, so it can vanish at
the horizon, if it has not vanished due to the shrinking of the $S^3$.
For large quark mass compared to the temperature (horizon size), the
$S^3$ shrinking will occur at some finite $u>u_H$, and the physics
will be similar to the zero temperature situation. However, for small quark
mass, the world--volume will vanish due to the shrinking of
the $S^1$ corresponding to the D7--branes going into the horizon. This
is new physics of the flavour sector.

The authors of ref.\cite{Babington:2003vm} explored some of the
physics of this situation (the dependence of the condensate and of the meson mass on the bare quark mass), and predicted that a phase transition
should occur when the topology of the probe D7--brane changes.  However, they were not able to explicitly see this transition because of poor data resolution in the transition region, coming from using UV boundary conditions on the scalar fields on the D7--brane world-volume.  The
origin of this phase transition, as we shall see, is as follows: The
generic behaviour of an allowed solution for $L(u)$ as in
equation~(\ref{eqt: L}), is not enough to determine whether the
behaviour corresponds to an $S^3$--vanishing D7--brane or an
$S^1$--vanishing D7--brane. The choices of branch of solutions have
different values of $c$, generically. In other words, for a given
value of $m$ there can be more than one value of $c$. There are
therefore two or more candidate solutions potentially controlling the
physics. The actual physical solution is the one which has the lowest
value for the D7--brane's free energy. The key point is that, at a
certain value of the mass, the lowest energy solution may suddenly
come from a different branch, and, as the corresponding value of the
condensate changes discontinuously in moving between branches, we find
that the system therefore undergoes a first order phase transition.  On the gauge theory side, we can imagine a similar situation occurring; two different branches of solution are competing, and the lowest energy branch is always picked.  
We are able to uncover this physics by doing a careful numerical analysis of the equations of motion for the probe dynamics on the gravity side of the AdS/CFT correspondence, which goes well beyond that carried out in previous papers. By using IR boundary conditions instead of UV boundary conditions, our analysis allows us to obtain vastly more data points, allowing us to complete the picture described above\footnote{In the process of writing this paper, similar work was published by Mateos et al.\cite{Mateos:2006nu}.  In addition, after this paper appeared, I. Kirsch pointed out to us some related work in his thesis \cite{Kirsch:2004km}.}. In addition,
we are able to study the meson spectrum in some detail, and we find
that the phase transition also manifests itself as a first order
discontinuity in that physics.
\section{The Probe Computation}
We begin by reviewing the physics of the D7--brane probe in the
AdS$_5$--Schwarzschild background solution\cite{Babington:2003vm}.  The metric is
given by:
\begin{equation} \label{eqt: our metric}
ds^2 = - \frac{f(u)}{R^2} dt^2 + \frac{R^2}{f(u)}  du^2 + \frac{u^2}{R^2} d\vec{x}\cdot d\vec{x} + R^2 d\theta^2 + R^2 \cos^2 \theta d \Omega_3^2  + R^2 \sin^2\theta d\phi^2\ ,
\end{equation}
where $\vec{x}$ is a three vector,
\begin{eqnarray}
f(u) &=& u^2 - \frac{b^4}{u^2} \nonumber \ ,
\end{eqnarray}
and the quantity $R^2$ is given by:
\begin{eqnarray}
R^2 &=& \sqrt{4 \pi g_s N_c} \alpha' \nonumber \ ,
\end{eqnarray}
where $g_s$ is the string coupling (which, with the inverse string
tension $\alpha^\prime$ sets for example, Newton's constant).  The quantity $b$ is related to the mass of the black-hole, $b^2=8 G_5 m_\mathrm{b.h.} / (3 \pi)$.  The
temperature of the black hole can be extracted using the standard
Euclidean continuation and requiring regularity at the horizon.
Doing this in the metric given by equation~(\ref{eqt: our metric}), we
find that $\beta^{-1}=b/\pi R^2$.  Therefore, by picking the value of $b$, we are choosing at which temperature we are holding the theory.  We choose to embed the D7--brane probe
transverse to $\theta$ and $\phi$.  In order to study the embeddings with the lowest value of the on--shell action (and hence the lowest free energy), we choose an ansatz of the form $\phi = 0$ and $\theta = \theta(u)$.  The asymptotic separation of the D3 and D7--branes is given by $L(u) = u \sin \theta$.  Given this particular choice of embedding, the worldvolume of the D7 brane is given by:
\begin{eqnarray}
 \sqrt{-g} &=& u^2 \cos^3 \theta(u) \sqrt{\det S^3} \sqrt{u^2 +\left(u^4-b^4 \right)\theta'(u)^2} \ ,
\end{eqnarray}
where $g$ is the determinant of the induced metric on the D7--brane given by the pull--back of the space--time metric $G_{\mu \nu}$.  We are interested in two particular cases.  First, there is the case where $u$ goes to~$b$,
which corresponds to the the D7--brane probe falling into the event horizon.  In the Euclidean section, this case corresponds to the shrinking of the $S^1$ of periodic time.
This corresponds to what in ref.\cite{Babington:2003vm} were called
condensate solutions.  Second, there is the case where $\theta$ goes to~$\pi/2$, which corresponds to the shrinking of the $S^3$.  This corresponds to what were called Karch-Katz--like solutions in
ref.~\cite{Babington:2003vm}.  It is this change in topology ($S^1$ versus $S^3$ shrinking) between the different solutions that will correspond to a phase transition.
The classical equation of motion for $\theta(u)$ is:
\begin{eqnarray} \label{eqt: eqt of motion}
\frac{d}{d u} \left( \frac{u^2\left(u^4-b^4 \right) \theta'(u) \cos^3 \left(\theta(u)\right)} {\sqrt{u^2 +\left(u^4-b^4 \right)\theta'(u)^2}} \right) + 3 u^2 \cos^2 \theta(u) \sin \theta(u)\sqrt{u^2 +\left(u^4-b^4 \right)\theta'(u)^2} &=& 0\ .
\end{eqnarray}
When $u$ goes to infinity and the background metric becomes asymptotically AdS$_5 \times S^5$, the equation of motion reduces to:
\begin{eqnarray}
\frac{d}{d u} \left(u^5 \theta'(u) \right) + 3 u^2 \theta(u) &=& 0,
\end{eqnarray}
which has solution:
\begin{eqnarray}
\theta(u) &=& \frac{1}{u} \left(m + \frac{c}{u^2} \right) \ .
\end{eqnarray}
These two terms are exactly the non--normalizable and normalizable terms corresponding to a dimension 3 operator ($\bar{\psi} \psi$) in the dual field theory with source $m$ and vacuum expectation value (vev) $c$.
\section{Embedding Solutions}
We solve equation \reef{eqt: eqt of motion} numerically using a shooting technique.  Shooting from infinity towards the horizon, physical solutions are those that have a finite value at the horizon.  This will only be accomplished for a particular $m$ and $c$ value from equation
(\ref{eqt: L}), which would have to be delicately chosen by hand.
Therefore, if we instead start from the horizon with a finite
solution, it will shoot towards the physical asymptotic solutions we
desire.  In order to be able to analyze the phase transition between
the condensate and Karch-Katz--like solutions, we shoot from the
horizon for the condensate solutions and from $\theta= \pi/2$ for the
Karch-Katz--like solutions.  This technique avoids having to correctly
choose the boundary conditions at infinity, which is a sensitive
procedure, allowing us to have many more data points to analyze the
phase transition.  We impose the boundary condition that, at
our starting point--the horizon, we have:
\begin{eqnarray} \label{eqt: bc}
\theta'(u) \big|_{S^1\to 0} &=&  \frac{3 b^2}{8}  \tan \theta(b) \nonumber \\
\theta'(u) \big|_{S^3 \to 0} &=& \infty
\end{eqnarray}
We argue that this is the physical boundary condition to take; the first is simply a result of taking the limit of $u \to b$ in the equation of motion in equation \reef{eqt: eqt of motion}, whereas the second is a result of requiring no conifold singularity as the $S^3$ shrinks to zero size \cite{Karch:2006bv}.

\begin{figure}[!ht]
\begin{center}
\includegraphics[angle=0,
width=0.55\textwidth]{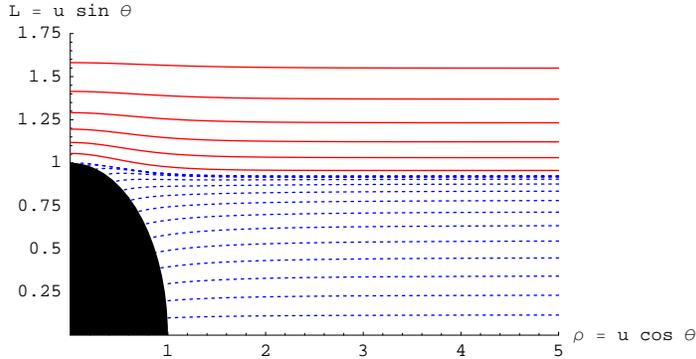}
\caption{Solutions for the D7--brane probe in the black hole background.}
\label{fig:solutionsinLandrho}
\end{center}
\end{figure}

We solve the equation of motion, equation~(\ref{eqt: eqt of motion}),
numerically using $b=1$ and $R=1$.  These numerical choices correspond to fixing the temperature and to measuring lengths in units of the radius of the AdS space; the latter condition also means we are choosing a particular relationship between the t'Hooft coupling and the dual quantities:
\begin{equation}
\lambda = g_{\text{YM}}^2 N_c = \frac{1}{2 \alpha^{\prime 2}} \ , \quad \beta^{-1} = \frac{1}{\pi} \ .
\end{equation}
Several D7--brane embedding solutions are shown in figure~\ref{fig:solutionsinLandrho}.  The red (solid) lines correspond to Karch-Katz--like solutions, and the blue (dashed) lines correspond to condensate solutions.  From each of these solutions, we can extrapolate the bare quark mass and quark condensate vev.
\section{The Phase Transition}
\begin{figure}[!ht]
\begin{center}
\subfigure[] {\includegraphics[angle=0,
width=0.45\textwidth]{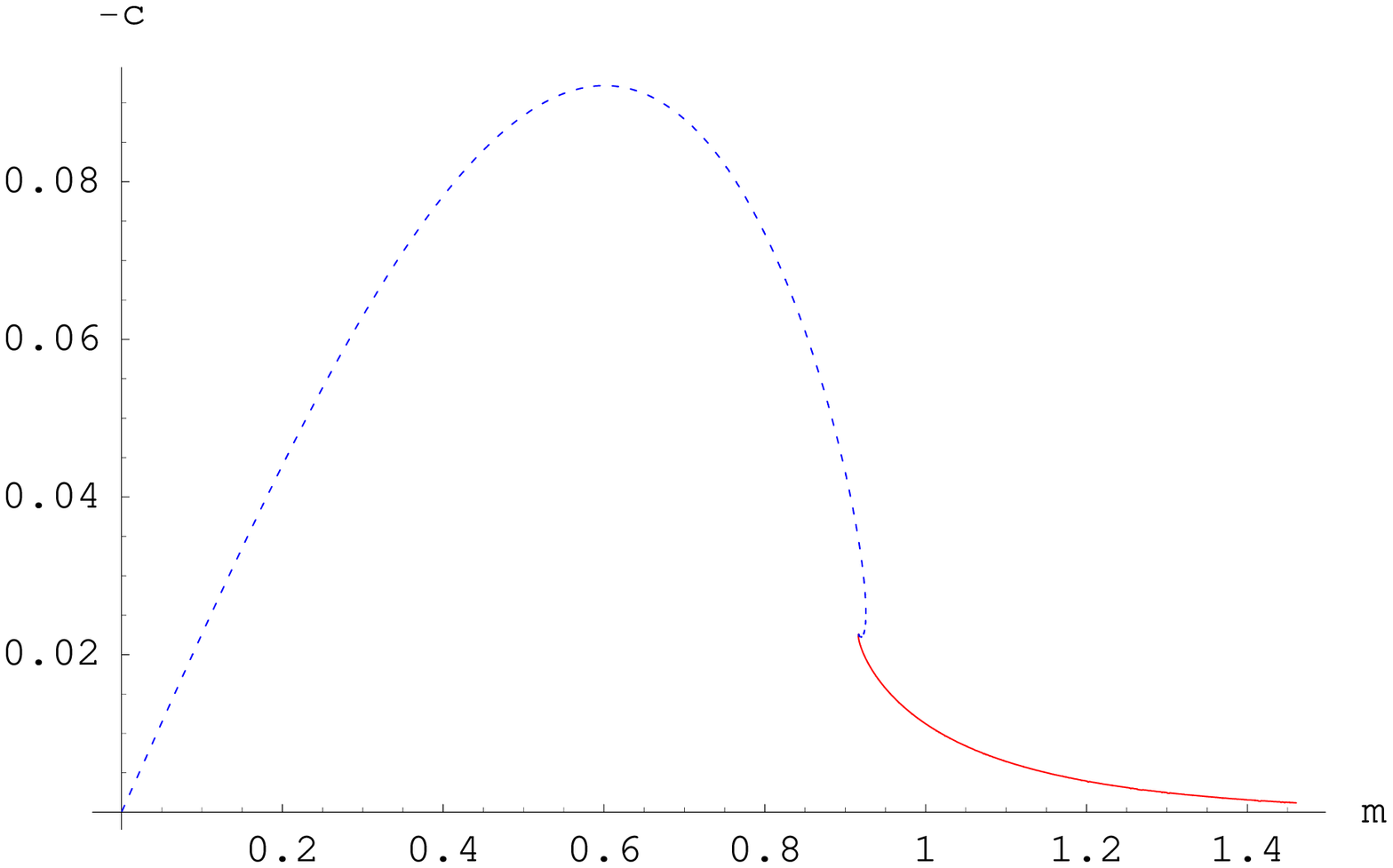} \label{fig: c vs m}}
\subfigure[] {\includegraphics[angle=0,
width=0.45\textwidth]{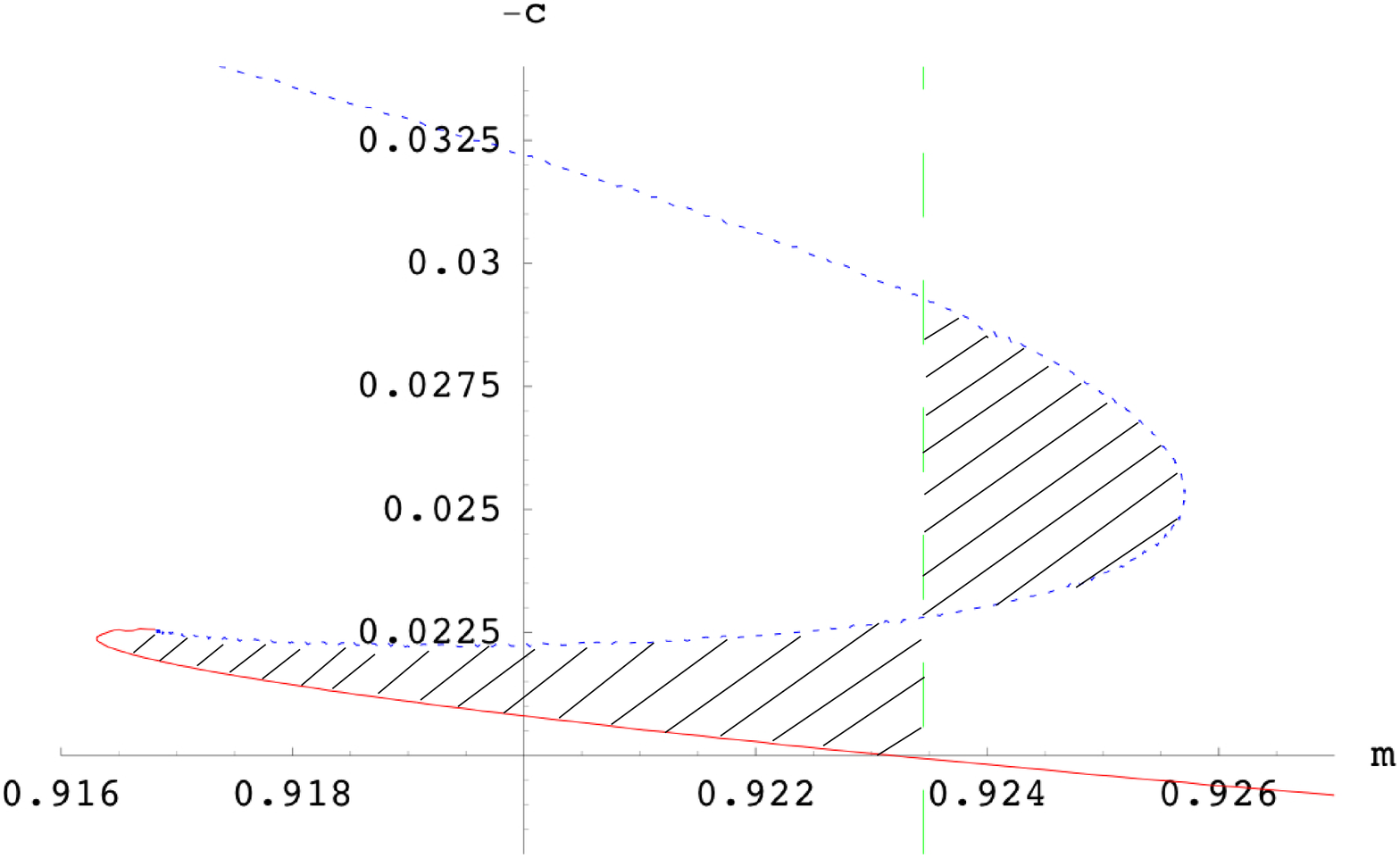} \label{fig: c vs m zoom}}
\subfigure[] {\includegraphics[angle=0,
width=0.45\textwidth]{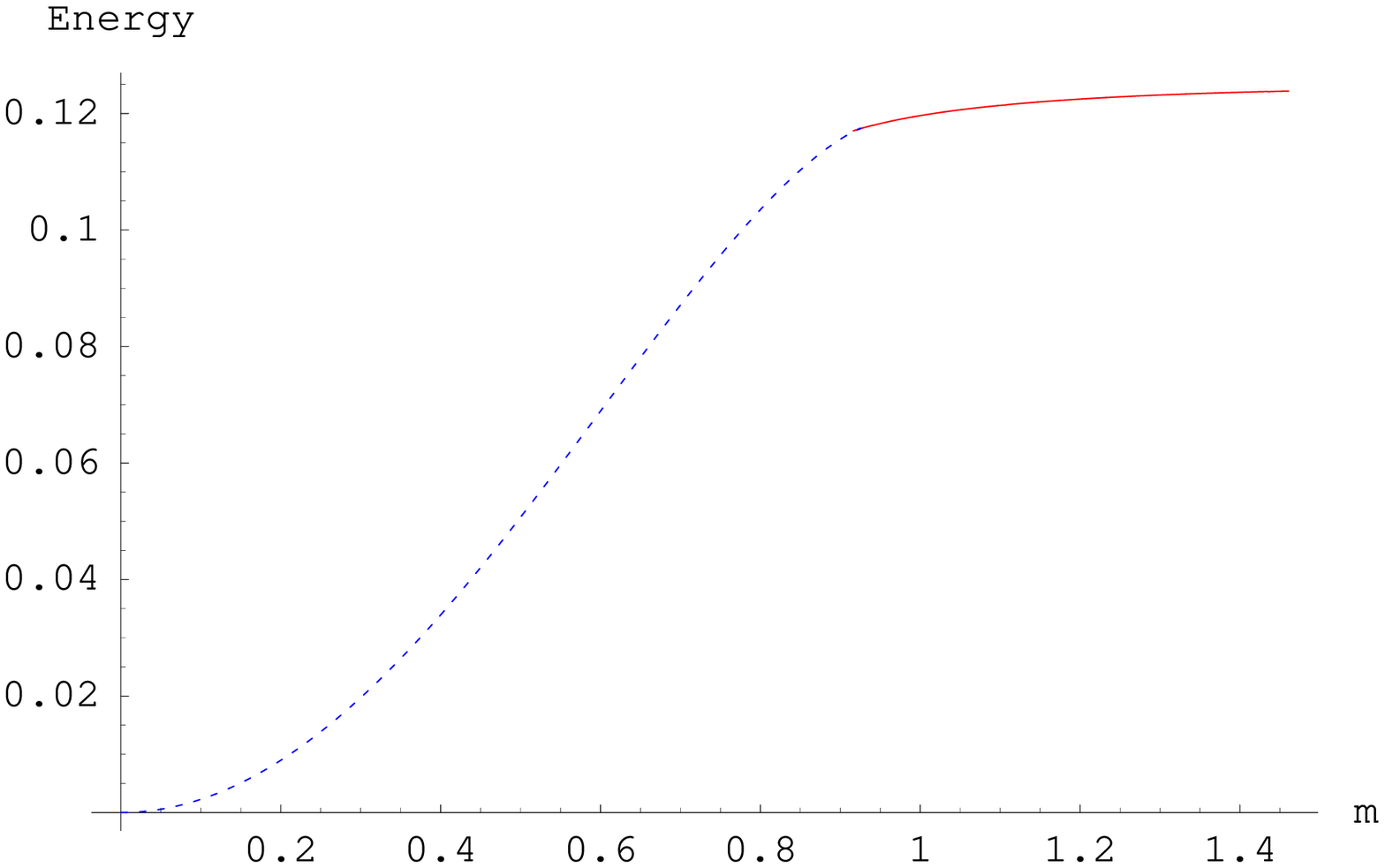} \label{fig: energy}}
\subfigure[] {\includegraphics[angle=0,
width=0.45\textwidth]{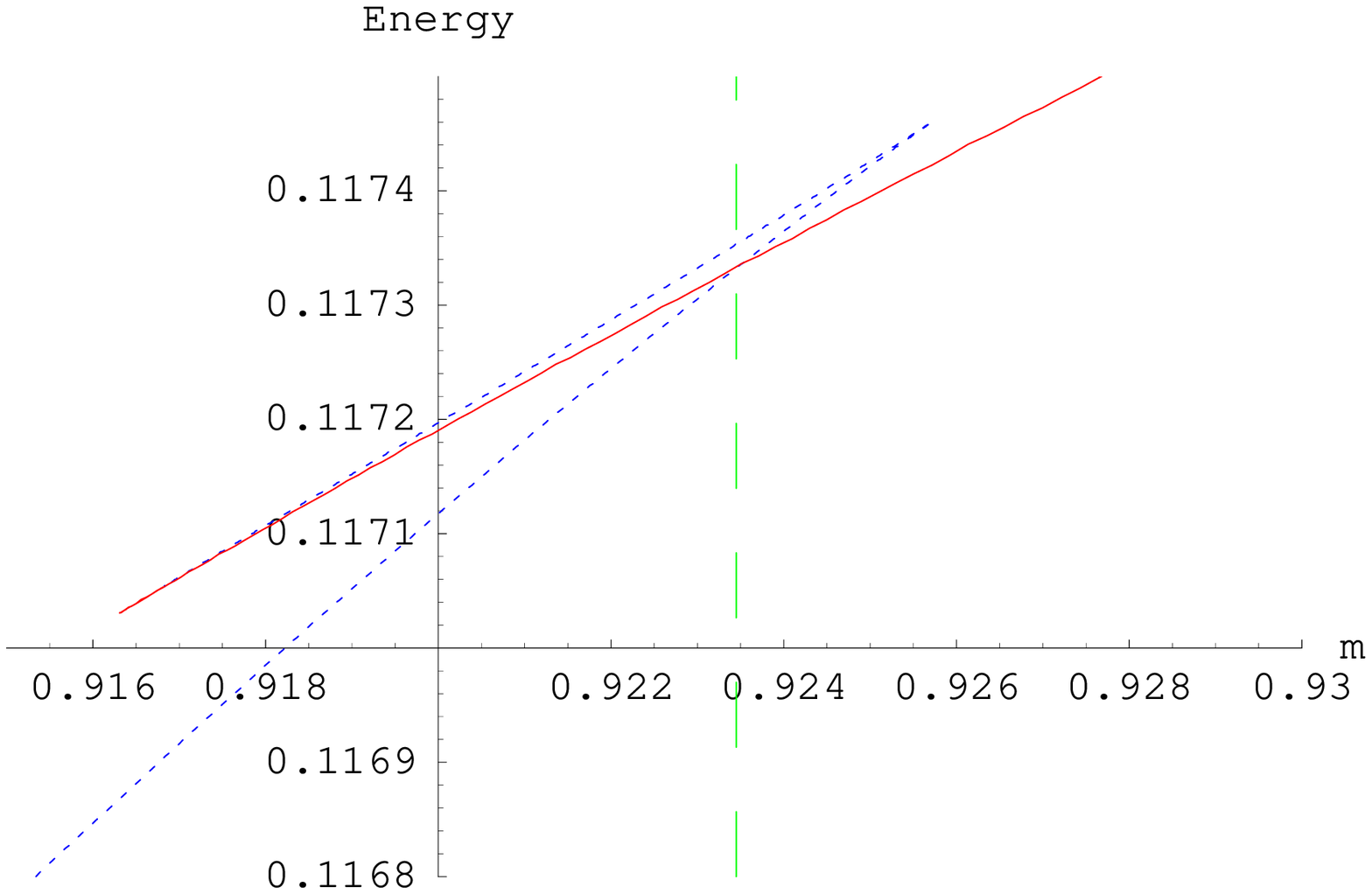} \label{fig: energy zoom}}
\caption{Probe free energy (in units of $2 \pi  \tau_7 N_f b/ R^2)$ and condensate vev at the phase transition. See text for more details.}
\end{center}
\end{figure}

We plot the $c$ values as a function of $m$ in figure~\ref{fig: c vs m}.  When enlarged, as shown in figure~\ref{fig: c vs m zoom}, we find, as anticipated in the introductory remarks, the multi--valuedness in $c$ for a given $m$. Physics will choose just one answer for $c$. There is therefore the possibility of a transition from one branch to another as one changes $m$.

In order to determine exactly where the transition takes place, we have to calculate the free energy of the D7--brane.  In the semi-classical limit that we are considering, the free energy is given by the the on--shell action times $\beta^{-1}$.  For our case, this is simply given by \cite{Kruczenski:2003uq}:
\begin{eqnarray}
\mathcal{F} &=& \beta^{-1} \tau_7 N_f \int d^4 x \ d\Omega_3 \ du \, \sqrt{-\det g}  \ ,
\end{eqnarray}
where here $N_f = 1$.  We calculate this integral numerically using our solutions for $\theta(u)$, and we plot the results for the energy in figure~\ref{fig: energy} after we regulate the result by subtracting off the energy from the $\theta=0$ solution.  There is again multivaluedness at the same value of $m$ as before, and we zoom in on this neighbourhood in figure~\ref{fig: energy zoom}.  If one follows the solutions of lowest energy for a given $m$, one can clearly see that there is a crossover from one branch to another as $m$ changes.  This was also observed in ref. \cite{Ghoroku:2005tf}.
Therefore, we find a first order phase transition ---at $m\approx 0.92345$--- where the condensate's vev jumps discontinuously.  This had been deduced on other grounds in ref. \cite{Apreda:2005yz}.  We show where the jump between curves occurs with the dashed green line in figures~\ref{fig: c vs m zoom} and~\ref{fig: energy zoom}.  We note that the shaded areas in figure~\ref{fig: c vs m zoom} are equal; this is to be expected since $c$ and $m$ are thermodynamically conjugate to each other, and the area of the graph has the interpretation as a free energy difference $d\mathcal{E} \sim - c \, dm$.
\section{Meson Spectrum}
\begin{figure}[ht!]
\begin{center}
\subfigure[] {\includegraphics[angle=0,
width=0.425\textwidth]{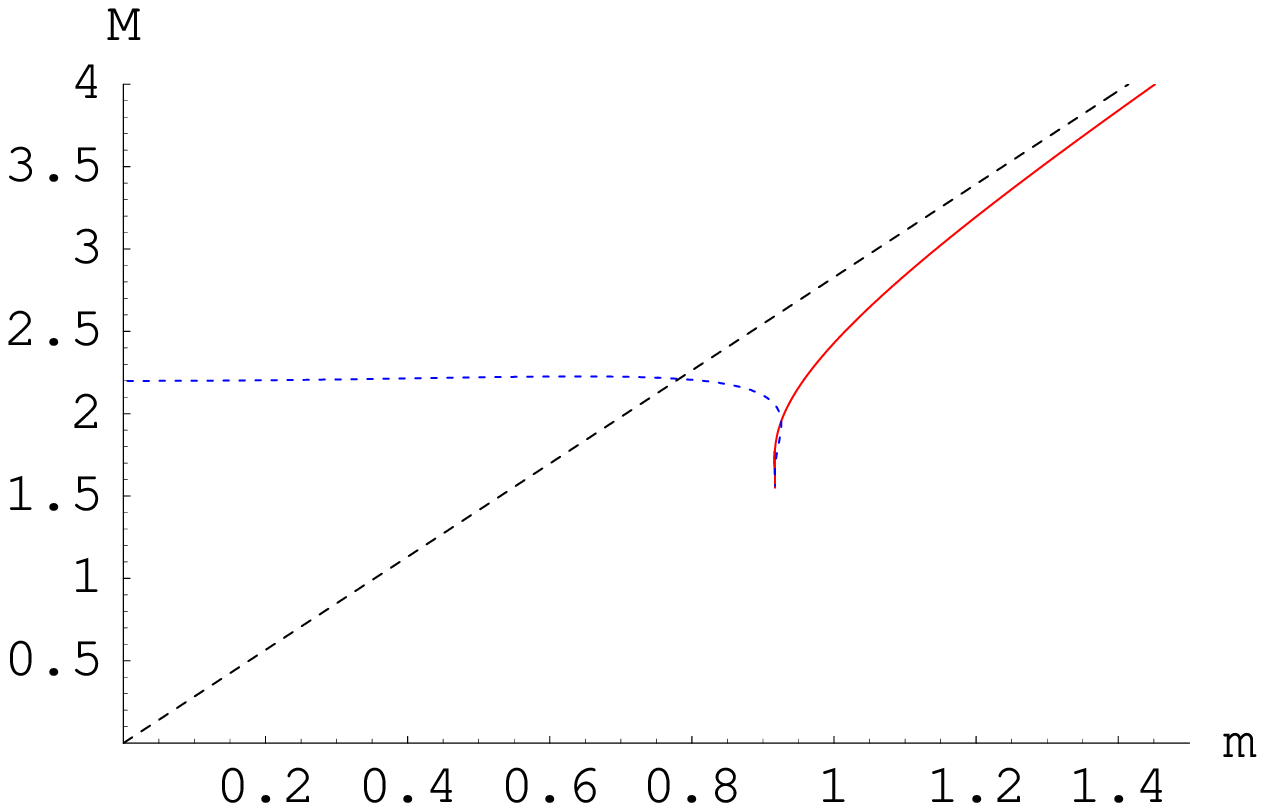} \label{fig: phi meson spectrum 4d}}
\subfigure[] {\includegraphics[angle=0,
width=0.425\textwidth]{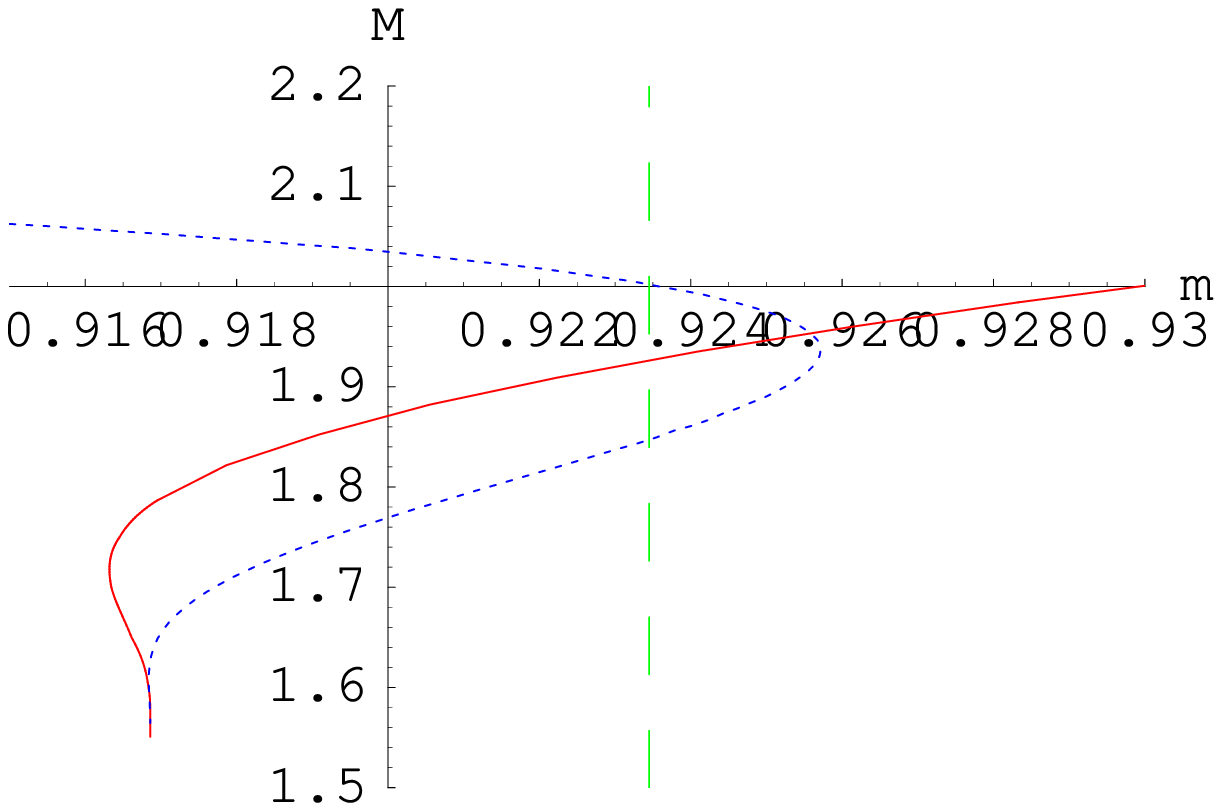} \label{fig: phi meson spectrum zoom 4d}}
\subfigure[] {\includegraphics[angle=0,
width=0.425\textwidth]{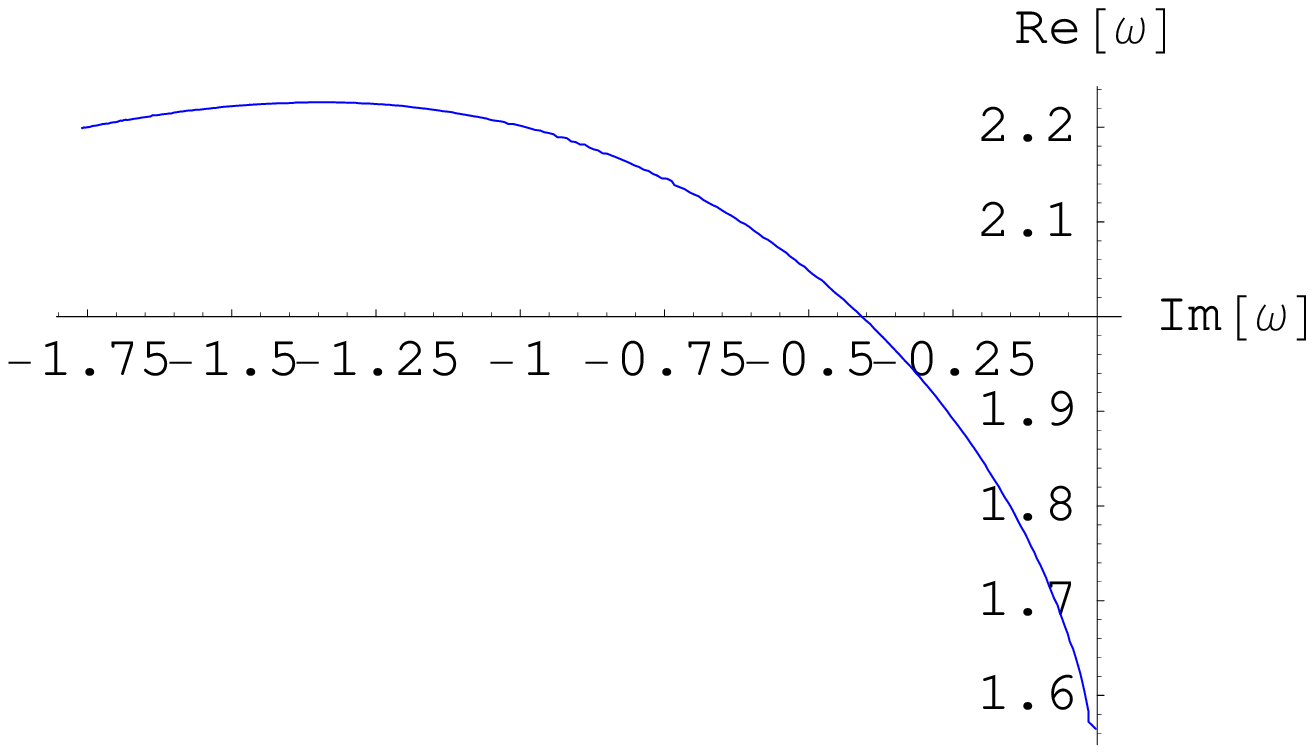} \label{fig:phi meson omega}}
\caption{Four--dimensional meson mass $M$, as a function of fundamental quark mass $m$, from fluctuations in $\phi$.  Only the first level of the discrete spectrum is shown.  The blue (dashed) curves in (a) and (b) correspond to the meson mass before the meson melts.}
\end{center}
\end{figure}

\begin{figure}[ht!]
\begin{center}
\subfigure[] {\includegraphics[angle=0,
width=0.425\textwidth]{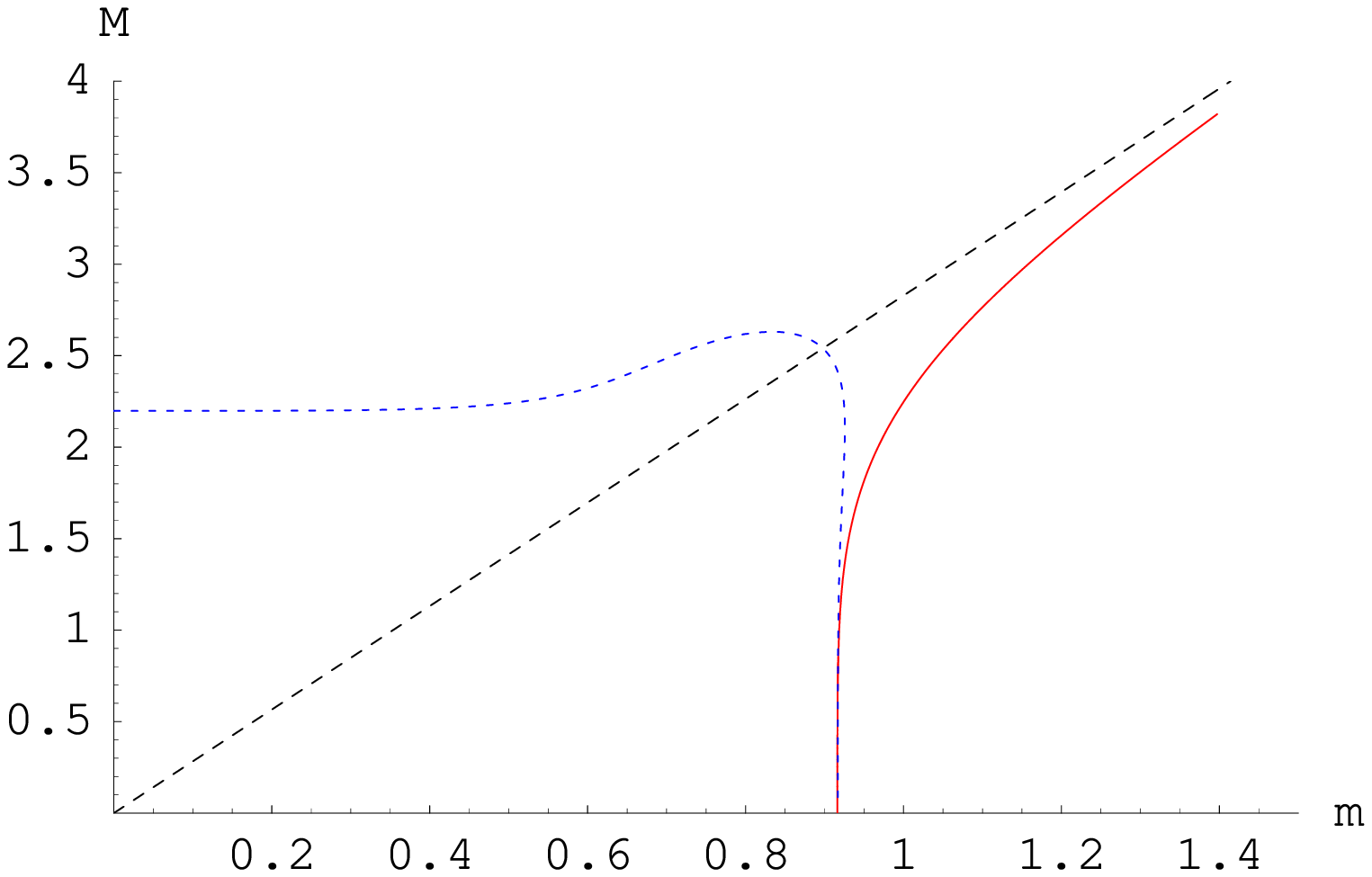} \label{fig: theta meson spectrum 4d}}
\subfigure[] {\includegraphics[angle=0,
width=0.425\textwidth]{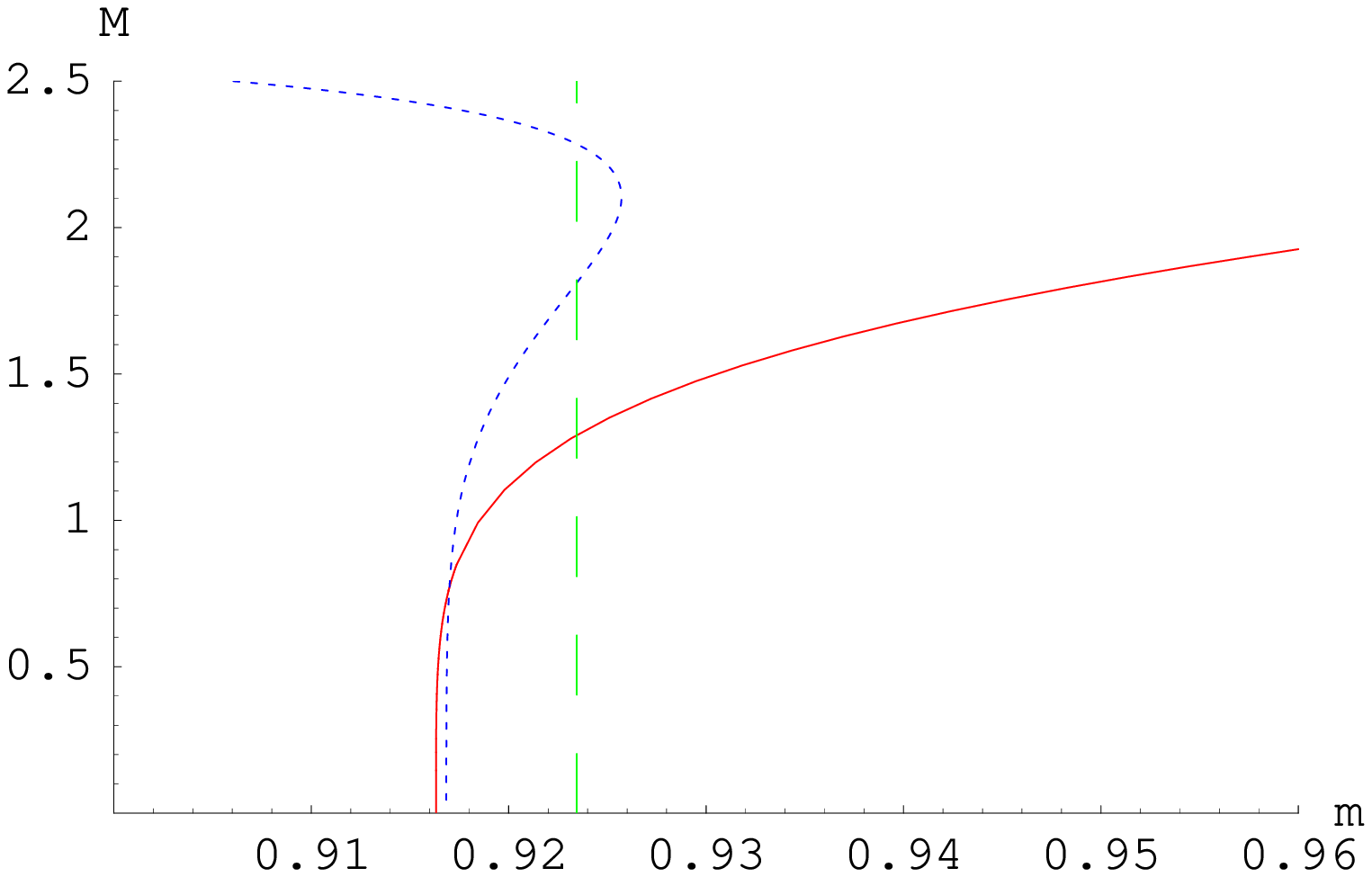} \label{fig: theta meson spectrum zoom 4d}}
\subfigure[] {\includegraphics[angle=0,
width=0.425\textwidth]{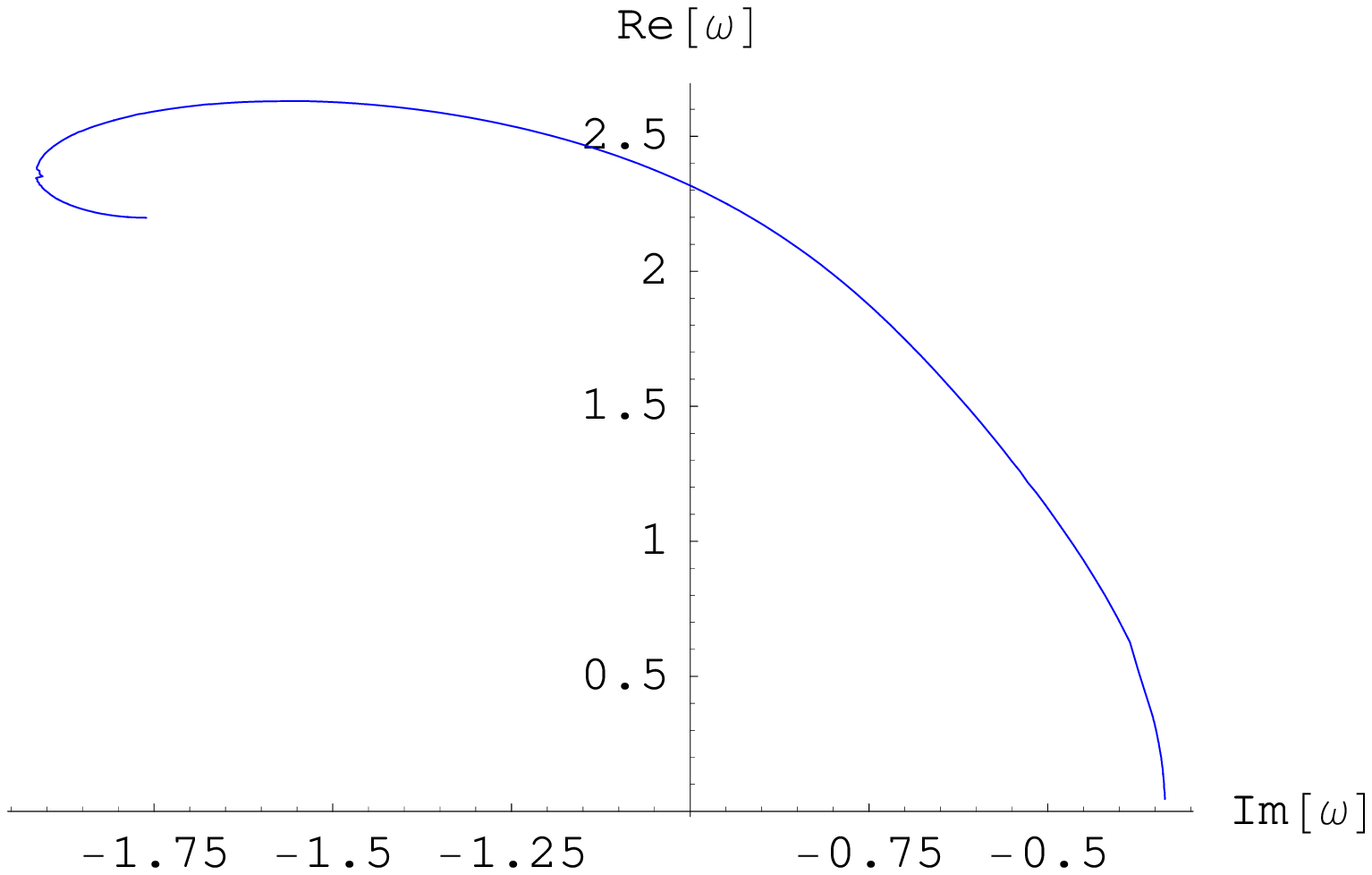}}
\caption{Four--dimensional meson mass $M$, as a function of fundamental quark mass $m$, from fluctuations in $\theta$.  Only the first level of the discrete spectrum is shown.  The blue (dashed) curves in (a) and (b) correspond to the meson mass before the meson melts.}
\end{center}
\end{figure}

\begin{figure}[ht!]
\begin{center}
\subfigure[] {\includegraphics[angle=0,
width=0.425\textwidth]{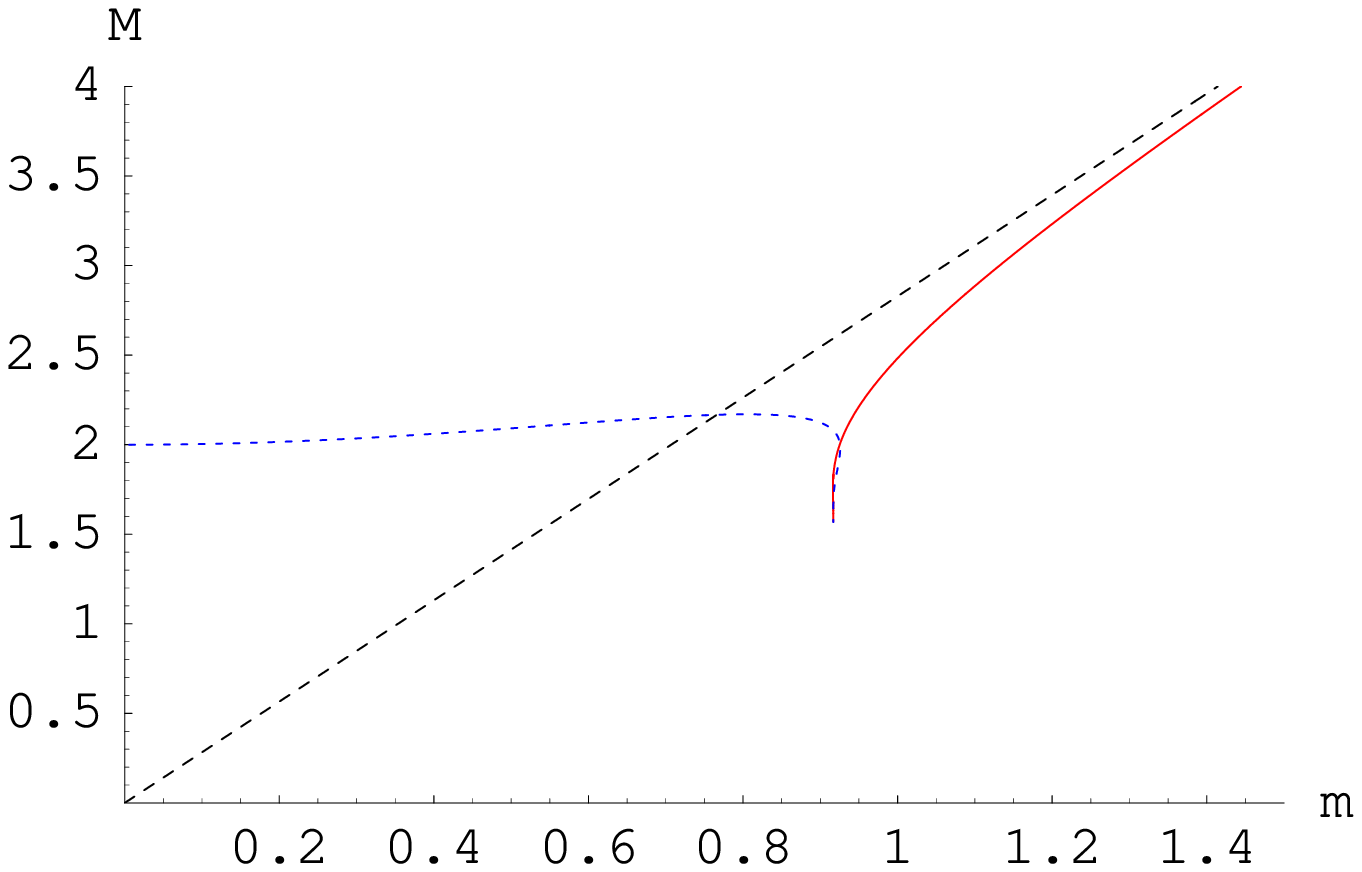} \label{fig: vector meson spectrum overall 4d}}
\subfigure[] {\includegraphics[angle=0,
width=0.425\textwidth]{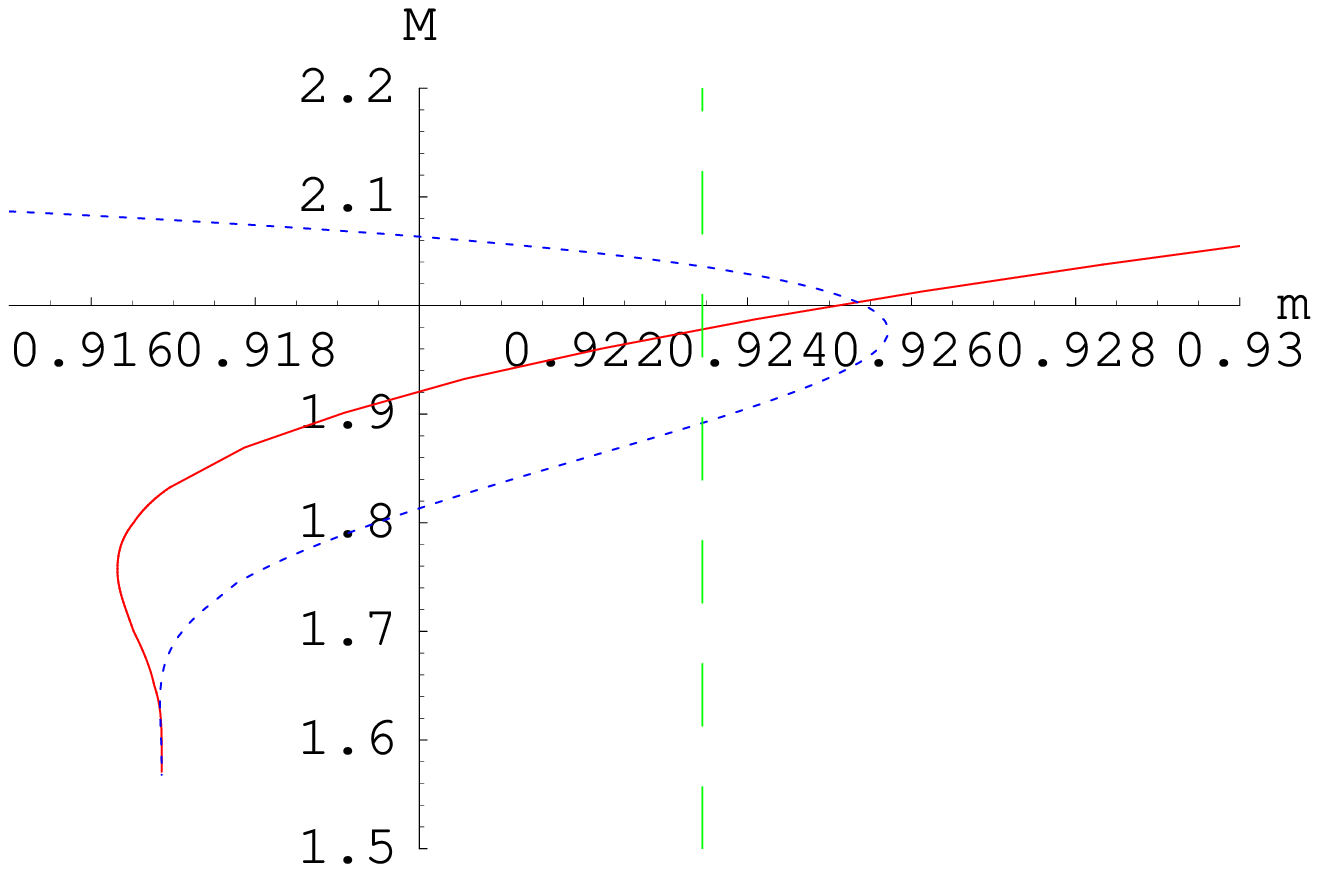} \label{fig: vector meson spectrum zoom 4d}}
\subfigure[] {\includegraphics[angle=0,
width=0.425\textwidth]{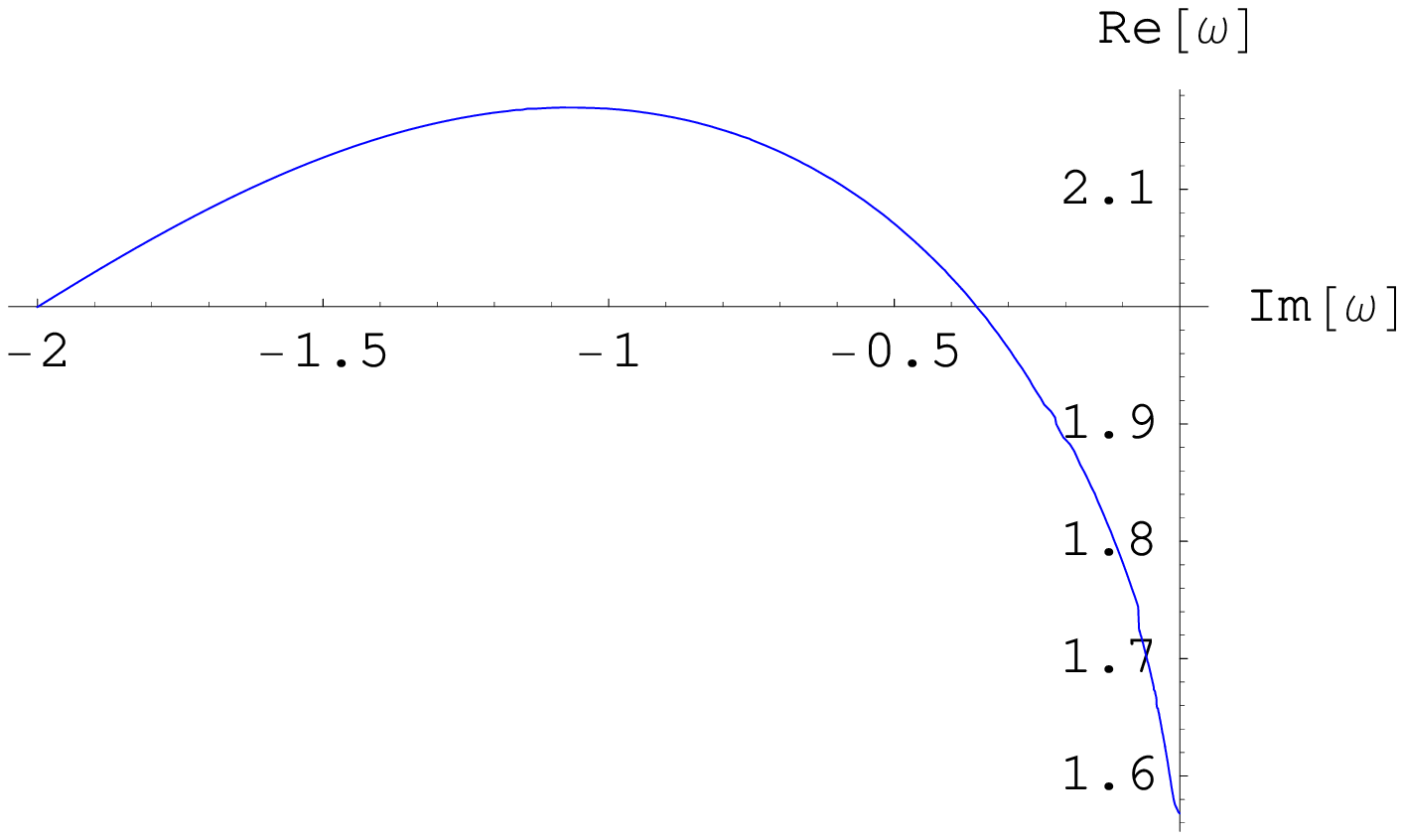}}
\caption{4--dimensional meson mass $M$, as a function of fundamental quark mass $m$, from fluctuations in $A_b$.  Only the first level of the discrete spectrum is shown.  The blue (dashed) curves in (a) and (b) correspond to the meson mass before the meson melts.}
\end{center}
\end{figure}

Next we study the meson spectrum. This can be read off from the physics of the fluctuations in~$\theta$ and $\phi$ about our classical solutions $\theta_{(0)}=\theta(u)$ and $\phi_{(0)}=0$ described above\cite{Karch:2002xe}, corresponding to scalar and pseudo scalar fields respectively in the gauge theory\cite{Kruczenski:2003be}.  As a reminder, in ref.~\cite{Kruczenski:2003be}, the exact meson spectrum  for the AdS$_5 \times S^5$ background was found to be given by:
\begin{eqnarray}
M(n,\ell) &=& \frac{2 m}{R^2} \sqrt{(n+\ell + 1)(n + \ell + 2)} \ ,
\end{eqnarray}
where $\ell$ labels the order of the spherical harmonic expansion, and $n$ is a positive integer that represents the order of the mode.  In our meson spectrum calculations, we have only considered the $n=\ell = 0$ solutions for the fluctuations in $\theta$ and $\phi$.
We begin by considering the four--dimensional mass of the meson corresponding to fluctuations in $\phi$.  By considering an ansatz of the form:
\begin{eqnarray}
 \phi(z,t)=0+\delta \phi(z,t) = f(z) e^{- i \omega t} \ ,
\end{eqnarray}
where $z = u^{-2}$ in terms of our previous coordinates, and by expanding to second order in $\delta \phi(u,t)$, we can calculate an equation of motion for $f(z)$.  In particular, we find that near the event horizon, the equation of motion becomes:
\begin{eqnarray}
f''(z) + \frac{f'(z)}{z-b^{-2}} + \frac{R^4 \omega^2}{16 b^2} \frac{f(z)}{\left(z-b^{-2}\right)^2} &=&0 \ .
\end{eqnarray}
This equation has solutions given by in--falling and out--going waves:
\begin{eqnarray}
f(z) &=& a (1 - z b^2 )^{i \frac{R^2 \omega}{4 b}} + b (1 - z b^2 )^{- i \frac{R^2 \omega}{4 b}} \ .
\end{eqnarray}
This analysis of the quasinormal mode behavior has been shown in refs.\cite{Starinets:2002br, Mateos:2007vn, Hoyos:2006gb}, and the (correct) boundary condition of in--falling waves corresponds to the fact that fluctuations about condensate solutions do not allow for mesons with a discrete mass spectrum\footnote{Refs.\cite{Mateos:2007vn,Hoyos:2006gb} appeared after the original version of this manuscript appeared, and were useful in refining this section's discussion.}.  The mesons ``melt'' away into the plasma, from the gauge theory perspective.  However, as was shown in ref.\cite{Hoyos:2006gb}, we can still find the mass of the meson (which we shall denote as $M$) for these fluctuations before they melt, in addition to the meson's lifetime ($\tau$) with the identification:
\begin{equation}
\mathrm{Re}[\omega] = M \ , \quad \mathrm{Im}[\omega] =  \left(2 \tau \right)^{-1} \ .
\end{equation}
The fluctuations about the Karch-Katz--like solutions have a discrete mass spectrum; therefore the mesons are stable, and $\omega$ will be purely real.  To search for the condensate fluctuations, we do a convenient field redefinition:
\begin{eqnarray}
f(z) &=& y(z) \left(1 - z b^2 \right)^{- i \frac{R^2 \omega}{4 b} } \ ,
\end{eqnarray}
such that our boundary condition is simply:
\begin{eqnarray} \label{eqt:bc1}
y(b^{-2}) &=& 1 \nonumber \\
y'(b^{-2}) &\sim& (1-z b^2 )^{-1} \big|_{z \to b^{-2}} \to \infty \ .
\end{eqnarray}
For the fluctuations about Karch-Katz--like solutions, we simply use the boundary conditions:
\begin{eqnarray} \label{eqt:bc2}
f(z)\big|_{\theta(z)=\pi/2} &=& \epsilon \nonumber \\
f'(z)\big|_{\theta(z)=\pi/2} &=& 0 \ ,
\end{eqnarray}
where $\epsilon$ is sufficiently small. (We used $\epsilon\sim 10^{-2}$ here.) In order to find the correct value of~$\omega$ for both sets of fluctuations, we note that in the limit that~$z$ approaches zero, where the background becomes asymptotically AdS$_5 \times S^5$, the field $f(z)$ must only be comprised of a normalizable term, {\it i.e.,} the coefficient in front of the non--normalizable term must be zero.
We show the first level of the discrete four--dimensional meson mass dependence on the bare quark mass in figures \ref{fig: phi meson spectrum 4d} and \ref{fig: phi meson spectrum zoom 4d} (zoomed).  In figure \ref{fig:phi meson omega} we show the behavior of the real and imaginary part of $\omega$.  \\
For large bare quark mass, our solutions match those of the AdS$_5\times S^5$ case discussed in ref.~\cite{Kruczenski:2003be} (shown as the black lines with long dashes).  We note that, for large bare quark mass, the meson mass is bounded from above by the AdS$_5\times S^5$ mass.  This is to be expected since the coupling in the finite temperature gauge theory goes as:
\begin{eqnarray}
\frac{1}{g_{\mathrm{YM}}^2}  &\sim& \left(1- \frac{b}{u} \right)^{-1/2} \frac{1}{g_{\mathrm{YM}}^2} \bigg|_{\beta=\infty}
\end{eqnarray}
In addition, we see that at zero bare quark mass, the meson has a finite mass.  This is simply the mass gap of the three--dimensional gauge theory:  In the Euclidean section, the four--dimensional gauge theory is compactified on an $S^1$, whose radius is set by the energy scale of the theory.  In this case, the energy scale is set by the bare quark mass, so at zero bare quark mass, the circle shrinks to zero size, making the gauge theory effectively three--dimensional.
\\
For fluctuations in the $\theta$ direction, we can follow a similar analysis as for the fluctuations in the $\phi$ direction.  We begin by searching for the 4--dimensional mass by taking an ansatz $\phi =0$
and $\theta = \theta_{(0)} + \delta \theta(z,t)$, where $\delta \theta(z,t)$ has the form:
\begin{equation}
\delta \theta(z, t) = f(z) e^{- i \omega t} \ .
\end{equation}
The equation of motion shows the same behavior as the $\phi$--meson, so we choose the same boundary condition as in equation \reef{eqt:bc1} for the condensate solutions.  For the Karch-Katz--like solutions, we use the boundary condition:
\begin{eqnarray} \label{eqt:bc3}
f(z)\big|_{\theta(z)=\pi/2} &=& \epsilon\ ,\nonumber \\
f'(z)\big|_{\theta(z)=\pi/2} &=& \infty \ ,
\end{eqnarray}
where $\epsilon$ is suitably small. (We chose $\epsilon\sim 10^{-2}$.) The spectrum is shown in figures \ref{fig: theta meson spectrum 4d} and \ref{fig: theta meson spectrum zoom 4d}.  In figure \ref{fig: theta meson spectrum zoom 4d}, the disconnect between the condensate and Karch-Katz--like solutions is due to there being tachyonic modes that extend beyond the zero meson mass \cite{Mateos:2007vn}, but we have not shown them in our plots. They occur well away from the physical regime.
\\
We may also consider fluctuations in the gauge field living on the D7--brane; such fluctuations would correspond to vector mesons in the gauge theory.  In order to study their behavior, we must begin by including the field strength term (from the world--volume gauge field) to the DBI action.  The action is given by:
\begin{equation}
S = - \tau_7 \int d^8 \xi \ \left( -\det \left(g_{a b} + 2\pi \alpha'  F_{a b} \right) \right)^{1/2} + \frac{\left(2 \pi \alpha' \right)^2}{2} \mu_7 \int P \left[ C_{(4)} \right] \wedge F \wedge F \ ,
\end{equation}
where $ P \left[ C_{(4)} \right] $ is the pull-back of the 4--form potential sourced by the $N_c$ D3--branes.  Our previous solutions assumed a solution $A_b = 0$ , so we will consider quadratic fluctuations about this solution.  Expanding this action to second order in $A_b$, we have:
\begin{equation}
S = -\tau_7 \int d^8 \xi \  \left[ \left(- \det\left(g_{a b} \right) \right)^{-1/2} \left( 1 - \frac{1}{4} (2 \pi \alpha' )^2   F_{a b} F^{b a} \right) - \frac{\left(2 \pi \alpha' \right)^2}{8} \frac{u^4}{R^4} F_{i j} F_{k l} \epsilon^{i j k l} \right]\ ,
\end{equation}
where the subscripts $a,b$ run over the world--volume coordinates and $i,j,k,l$ run over the world--volume coordinates transverse to the D3--brane.  We have dropped the first order term since it will just give the classical equation of motion for $A_b$.  Since we are only interested in the components of the gauge field along the D3--brane worldvolume, the Wess--Zemino term will not contribute to the equations of motion that we are interested in.  Therefore, the equation of motion for the quadratic fluctuations is given by:
\begin{equation}
 \partial_a \left(\sqrt{- g} F^{a \mu} \right) = 0 \ .
\end{equation}
To compute the 4--dimensional mass of the vector meson, we consider an ansatz of the form:
\begin{equation}
A_\mu = V_\mu(z) e^{-i \omega t} \ .
\end{equation}
We want to consider the case where the condition $k_\mu A^\mu = 0$ still holds, which requires us to set $A_t = 0$ since we are working in the rest frame of the vector meson.  The equations of motion for the three $V_{x^i}$'s are identical, so we only need to solve for one of them.  The equations of motion show the same behavior as the $\phi$--mesons, so we follow the same procedure as before with the same boundary conditions as in equations \reef{eqt:bc1} and \reef{eqt:bc2} .  The spectrum is shown in figures \ref{fig: vector meson spectrum overall 4d} and \ref{fig: vector meson spectrum zoom 4d}.  We see again that the mass is bounded from above by the AdS$_5\times S^5$ mass (shown as the black lines with long dashes).
\\
\section{Conclusion}
\label{sec:conclusion}
We expect that, while the details of this construction will not persist
in a ``realistic'' QCD string dual, the phase transition itself
represents strongly coupled dynamics that may well persist as part of
the full story of the QCD phase diagram. The transition is exciting in
itself, of course (particularly since its dual involves a change of
topology in the D7--brane world--volume), but much further work is
needed on several questions. For example, the robustness of the phase
transition against $1/{N_c}$ corrections would be interesting to
study. It would also be
interesting to see if the back--reaction can be included, allowing us to
study $N_f\sim N_c$ and follow the phase structure to this regime.
Although there are complications involving conical deficits in
otherwise flat directions when dealing with the back--reacted geometry
in the case of a D3--D7 system, some progress has been made in studying
the back-reacted geometry of other more stable Dp--D(p+4) systems
\cite{Grana:2001xn, Bertolini:2001qa, Cherkis:2002ir, Nastase:2003dd,
  Burrington:2004id, Erdmenger:2004dk, Casero:2006pt}.  However, many
of these calculations have not  considered
black hole or other finite temperature configurations.  These, and several
other questions, are exciting matters for further study.

\section*{Acknowledgments}
This research is supported by the US Department of Energy. We have benefitted from discussions with Robert C. Myers and David Mateos.

\providecommand{\href}[2]{#2}\begingroup\raggedright\endgroup
\end{document}